\documentclass[12pt,preprint]{aastex}

\newcommand{\fad}{Faraday~Discuss.}   
\newcommand{\jmsp}{J.~Mol.~Spectr.}   
\newcommand{\jpca}{J.~Phys.~Chem.~A}  
\newcommand{\jpcs}{J.~Phys.:~Conf.~Ser.}  

\shorttitle{Discovery of Interstellar Propylene}
\shortauthors{Marcelino et al.}

\begin{document}

\title{Discovery of Interstellar Propylene (CH$_2$CHCH$_3$):\\
Missing Links in Interstellar Gas--Phase Chemistry}

\author{N. Marcelino\altaffilmark{1,4},
J. Cernicharo\altaffilmark{1},
M. Ag\'undez\altaffilmark{1},
E. Roueff\altaffilmark{2},
M. Gerin\altaffilmark{3}, 
J. Mart\'{\i}n-Pintado\altaffilmark{1},
R. Mauersberger\altaffilmark{4},
and
C. Thum\altaffilmark{5}}

\altaffiltext{1}{Departamento de Astrof\'{\i}sica Molecular e
Infrarroja, Instituto de Estructura de la Materia, CSIC, Serrano
121, 28006 Madrid, Spain; nuria@damir.iem.csic.es,
cerni@damir.iem.csic.es, marce@damir.iem.csic.es,
martin@damir.iem.csic.es}

\altaffiltext{2}{LUTH, Observatoire de Paris, Section de Meudon,
Place Jules Janssen, 92195 Meudon, France;
evelyne.roueff@obspm.fr}

\altaffiltext{3}{LERMA, UMR 8112, CNRS, Observatoire de Paris and
\'Ecole Normale Sup\'erieure, 24 Rue Lhomond, 75231 Paris Cedex
05, France; gerin@lra.ens.fr}

\altaffiltext{4}{IRAM, Av. Divina Pastora, 7, 18012 Granada, Spain; 
mauers@iram.es}

\altaffiltext{5}{IRAM, 300 rue de la Piscine, Domaine 
Universitaire 38406 Saint Martin d'H\`eres, France; thum@iram.fr}

\begin{abstract}
We report the discovery of propylene (also called propene, CH$_2$CHCH$_3$)
with the IRAM 30-m radio telescope toward the dark cloud \objectname{TMC-1}. 
Propylene is the most saturated hydrocarbon ever detected
in space through radio astronomical techniques. In spite of its weak dipole
moment, 6 doublets ($A$ and $E$ species) plus another line from the $A$ species 
have been observed with main beam temperatures above
20 mK. The derived total column density of propylene is 4$\times10^{13}$ cm$^{-2}$,
which corresponds to an abundance relative to H$_2$ of 4$\times10^{-9}$, i.e., 
comparable to that of other well known and abundant hydrocarbons in this cloud, 
such as c-C$_3$H$_2$. Although this isomer of C$_3$H$_6$ could play an important 
role in interstellar chemistry, it has been ignored by previous chemical models of 
dark clouds as there seems to be no obvious formation pathway in gas phase.
The discovery of this species in a dark cloud
indicates that a thorough analysis of the completeness of gas phase
chemistry has to be done.

\end{abstract}

\keywords{astrochemistry --- line: identification --- ISM: abundances --- 
ISM: clouds --- ISM: molecules}

\section{Introduction}
Cold ($T_{\rm kin}\sim$10 K), moderately dense ($n(\rm H_2)\sim10^4$ cm$^{-3}$) 
dark clouds are the sites of low mass star formation.
They exhibit a complex gas-phase ion-neutral chemistry leading to the 
formation of a large variety of molecules. The lack of internal heating 
sources and violent physical processes, like shocks, make these dense and
quiescent cores the best sites to explore and to model interstellar
gas-phase chemistry and molecular depletion into the dust grain surfaces.
The dense filament TMC-1, embedded in the Heiles Cloud 2
\citep[see, e.g.,][]{Cer87} in the Taurus region, is one of the richest 
molecular sources, with a large wealth of organic species such as 
unsaturated carbon chain radicals (C$_n$H) and cyanopolyynes (HC$_{2n+1}$N). 
Whereas a common and well adopted view in the literature is to consider 
this region as dense, cold and quiescent (as indicated by the extremelly 
narrow lines), some authors suggest that the 
presence of shocks could induce explosive injection of grain mantles 
and contribute to the formation of large carbon molecules such as 
methylcyanodiacetilene \citep{Mar00,Sny06}.

The history of molecular astronomy shows that interstellar molecules 
can be detected by specific searches, when frequencies are known 
and one has hints that a certain molecule is abundant. The most 
unexpected discoveries, however, resulted from spectral line 
surveys or were serendipitous.
To our knowledge there are only two spectral line surveys of dark 
clouds, both in the direction of TMC-1. One was carried out using 
the Nobeyama radio telescope \citep{Ohi98,Kai04} in the 8.8-50 GHz 
domain, with 414 detected lines arising from 38 different molecules, 
11 of them unknown prior to the survey.
Only two of the 38 species are diatomic (CS and SO), four are triatomic
and the rest are polyatomic, the longest one having 11 atoms (HC$_9$N).
The other survey was performed at lower frequencies, with the Arecibo 
radio telescope \citep{Kal04}, covering the range 4--6 GHz 
besides selected frequencies in the 8--10 GHz interval. 
They detected 29 lines arising from carbon-chain species such as C$_4$H,
H$_2$C$_4$, HC$_3$N, HC$_5$N, HC$_7$N, HC$_9$N, C$_2$S and C$_3$S.

Apart from CO and NO, all observed molecules in dark clouds 
usually have large dipole moments and most of them have rather 
low abundances (a few 10$^{-10}$). 
From this, one would naively expect that only polyatomic molecules with 
a large dipole moment should be detectable.
Due to the low kinetic temperature of these clouds, heavy molecules will
have their emission peak in the centimeter and millimeter wavelength 
domains.
As a result of a 3\,mm spectral line survey of several dark cloud cores 
(Marcelino et al. in prep.) and follow-up observations performed thereafter, 
we report the first detection of interstellar propylene (CH$_2$CHCH$_3$). 
This unexpected, nearly saturated hydrocarbon was detected toward TMC-1. 
This species, which is naively expected in hot cores rather than in dark 
clouds, has so far not been considered in any chemical model of TMC-1 
in spite of having an abundance similar to that of c-C$_3$H$_2$.

\section{Observations}
The observations were carried out with the IRAM 30-meter 
telescope\footnote{IRAM is supported by INSU/CNRS (France), 
MPG (Germany) and IGN (Spain).} near Granada (Spain) 
Two SIS receivers sensitive in the 3\,mm band were used 
simultaneously at the same frequency. Both receivers were 
tuned in single sideband mode with image rejections of 
$\sim$20 dB. System temperatures were between 
110--150 K during all observing periods. We used frequency 
switching mode with a frequency throw of 1.8 MHz (survey 
observations), while a throw of 7.2 MHz was used to obtain the 
new spectra outside the survey range in order to remove 
instrumental ripples and to improve baselines \citep[see,][]{Thu95}. 
The backend was the VESPA autocorrelator, with 40 kHz of spectral 
resolution ($\sim$0.13 km\,s$^{-1}$), providing a maximum effective 
bandwidth of 315 MHz. Intensity calibration was performed using two 
absorbers at different temperatures, and the atmospheric opacity 
was obtained from measurements of the sky emissivity and using 
the ATM code \citep{Cer85}. On-source integration times were 
typically $\sim$3 hours. Pointing and focus were checked on 
strong and nearby sources every 1.5 and 3 hours respectively.

The temperature scale is main-beam brightness temperature ($T_{\rm MB}$).
At the observed frequencies the beamwidth of the antenna is 24--29$''$ 
and the main beam efficiency is 0.78--0.75.
The observed lines are shown in Fig.~\ref{fig:fig_lines} and the 
line parameters are given in Table~\ref{tab:lines}.

\section{Results}
\begin{figure}
\centering
\includegraphics[scale=.25]{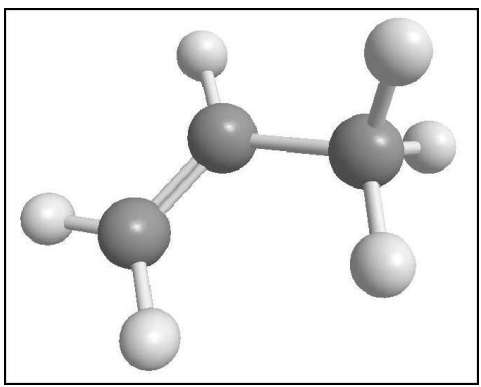}
\caption{Molecular structure of Propylene. Spheres represent carbon (dark grey) and hydrogen (light grey) atoms.
\label{fig:fig_molec}}
\end{figure}

During the 3\,mm line survey of TMC-1, which was obtained in several
observing runs between 2003 and 2007,
we detected a pair of lines at 86.650 GHz with similar intensities and
separated by 3 km\,s$^{-1}$, towards the cyanopolyyne peak (CP) of this
source [R.A.(J2000.0)=04$^{\rm h}$ 41$^{\rm m}$ 41.9$^{\rm s}$, 
Decl.(J2000.0)=+25$^\circ$ 41$'$ 27.1$''$].
These lines appear neither in the JPL nor in the CDMS catalogs 
\citep{Pic98,Mul01,Mul05}. Using the spectral line catalog
developed and maintained at the DAMIR-IEM-CSIC \citep{Cer00},
we tentatively identified these two lines as one doublet arising from the
$(5_{05}-4_{04})$ rotational transition ($A$ and $E$ species) of propylene
(also called propene, CH$_2$CHCH$_3$) in the survey data of TMC-1.
This molecule, whose structure is depicted in Fig.~\ref{fig:fig_molec}, 
is an internal rotor with a moderately high barrier against the torsion 
of the methyl group. This internal rotation produces a splitting of each 
rotational line by a value in the range of 0.3 MHz to 4 MHz, each member 
of the doublet belonging to the $A$ and $E$ species. Its rotational spectrum 
has been measured in the laboratory between 8--245 GHz \citep{Wlo94} and up 
to 412 GHz \citep{Pea94}, which allows 
accurate frequency predictions within the frequency domain of our search. 
It has a small dipole moment, $\mu_a$=0.36 D, $\mu_b$=0.05 D \citep{Lid57}.

\begin{figure}
\includegraphics[scale=.65]{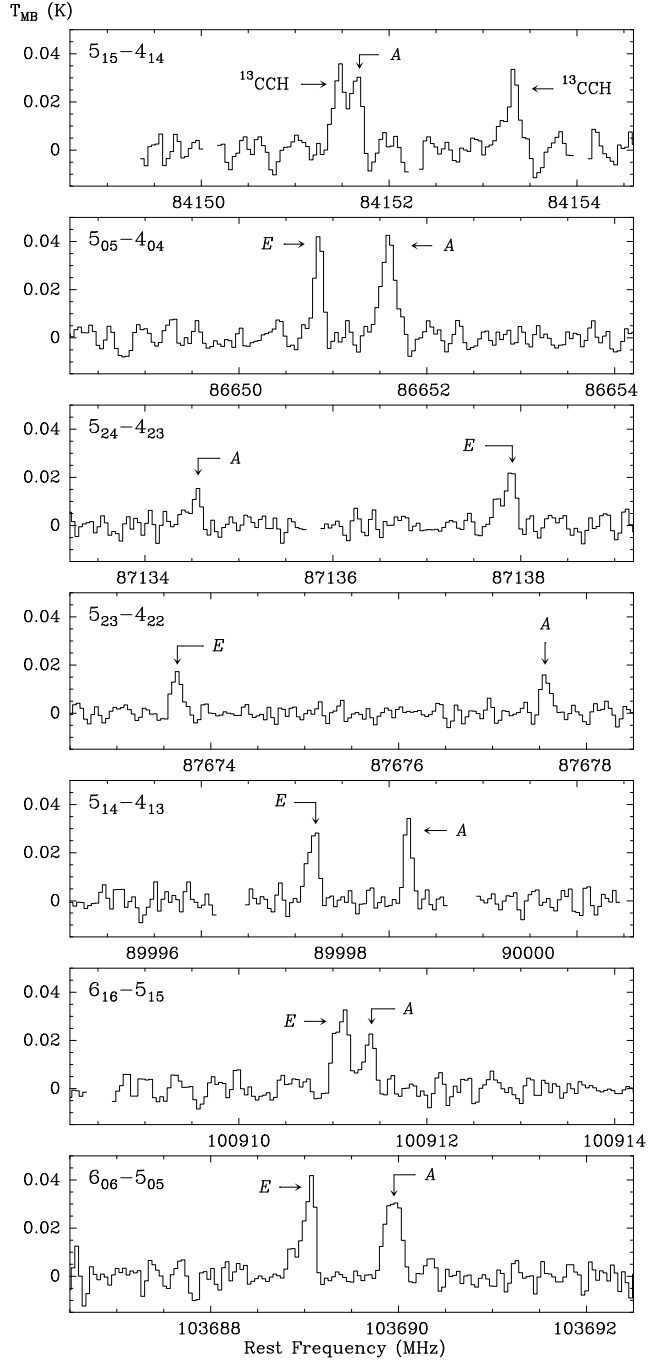}
\caption{Observed transitions arising from both $A$ and $E$
species of CH$_2$CHCH$_3$. Rest frequency is computed for
an $v_{LSR}$ = 5.6 km s$^{-1}$. Note on the top panel that only the
$A$ species was detected (see text). \label{fig:fig_lines}}
\end{figure}

Encouraged by this tentative identification, we searched for other 
transitions of this molecule in the frequency range covered by our survey 
(85.9--93 GHz). The fact that the lines must appear as a doublet separated 
by 0.3--4 MHz, and having similar intensities helped us in the search. 
Using the mentioned catalog, which contains only complete predictions for 
the $A$ species, together with the frequencies measured in the laboratory 
for the $E$ species \citep{Pea94}, three more doublets at 87.1 GHz, 87.7 
GHz and 90.0 GHz were identified at the 3--4 $\sigma$ level.
Director's Discretionary Time at the 30-m telescope was used to confirm 
the identification of propylene in April and May 2007: We re-observed 
the latter three mentioned doublets and searched for three
additional doublets lying outside the frequency range of the
survey, at 84.1 GHz, 100.9 GHz and 103.7 GHz. We detected all the
lines searched for, except for the $E$ component of the
5$_{15}$-4$_{14}$ transition at 84.1 GHz, for which the frequency
is not known, since this transition was not measured in the 
laboratory by \citet{Pea94}. However, since the separation in 
frequency of each doublet is small, the $E$ line could be hidden by one
hyperfine component of the $N = 1-0$ transition of $^{13}$CCH at
84151.5 MHz (see top panel in Fig.~\ref{fig:fig_lines}). In fact,
this hyperfine component of $^{13}$CCH has a line strength twice
less than that at 84153.4 MHz (also shown in the top panel of
Fig.~\ref{fig:fig_lines}). Therefore, with a total of 13 observed
lines, the first detection of interstellar propylene is
definitively confirmed in TMC-1.

Line parameters were obtained from gaussian fits
using the GILDAS package\footnote{http://www.iram.fr/IRAMFR/GILDAS} 
and are given in Table~\ref{tab:lines}.
All the observed lines are centered at a systemic velocity around 
5.6 km s$^{-1}$, which is in agreement with that derived from 
other molecules in TMC-1 \citep{Kai04}. However, several 
lines show a weaker second peak at a higher velocity, that could 
arise from the emission of a different component of the cloud, 
at 5.9 or 6.1 km s$^{-1}$ \citep{Pen98,Dic01}.
Line widths range from 0.3 to 0.6 km\,s$^{-1}$, being sometimes 
wider than the expected value of $\sim$0.4 km\,s$^{-1}$. 
The larger widths could imply again the emission from another 
component of the cloud.

Collisional rates are not available
for this molecule. Hence we have used a standard rotational diagram, 
including both species, to derive its column density and the rotational 
temperature, $T_{\rm rot}$, of the energy levels. 
We have obtained $T_{\rm rot}$ = 9.5$\pm$2.0 K and
$N(A$-CH$_2$CHCH$_3) = N(E$-CH$_2$CHCH$_3$) = 
(2.0$\pm$0.8) $\times$ 10$^{13}$ cm$^{-2}$,
i.e., a total column density for propylene of 
(4.0$\pm$1.5) $\times$ 10$^{13}$ cm$^{-2}$.
The rotational temperature is similar, within the errors, to that found
for other species in TMC-1 \citep{Ohi98,Fos01,Kal04}. 
However, the column density of propylene is particularly high for
a saturated species and for a molecule having 9 atoms 
(compare Table~\ref{tab:dens}). It
is similar to that of $c$-C$_3$H$_2$ and to that of CH$_3$CCH, which 
also contain the methyl group and has a rather low dipole moment 
(see Table~\ref{tab:dens}).
It is worth to note that for some transitions the $E$ species shows 
a slightly higher intensity than the $A$ species (see 
Fig.~\ref{fig:fig_lines} and Table~\ref{tab:lines}), then 
the column density could be higher for the $E$ species.
In fact, performing separate rotational diagrams for the two 
species, the derived rotational temperature for the $E$ species, 
$T_{rot}$ = 12 K, is slightly higher than for the $A$, $T_{rot}$ = 8 K.
The total column density, however, do not differ much from the 
value obtained with a single rotational diagram. Hence, within 
experimental accuracy both species seem to have the same abundance.

\section{Discussion}

Propylene, a rather simple hydrocarbon, has escaped detection because 
apparently nobody ever considered a directed search worthwhile, and also
because of its small dipole moment, 10 times
smaller than that of other three carbon molecules (see 
Table~\ref{tab:dens}). \citet{Pea94} already
suggested that this molecule could be detected in space. However,
as this hydrocarbon is nearly saturated, they pointed out that the
best place to search for it could be hot cores in high-mass star
forming regions. Prior to this detection, we had already searched for
propylene in the spectral line survey of Orion (Tercero and
Cernicharo, in preparation), but no lines were found within a 
detection limit of 50 mK.
We have checked the U-lines in previous line surveys of Orion
and Sgr B2 at different frequency bands
\citep{Sut85,Cum86,Tur89,Sch97,Num98,Fri04}, and also the 
unidentified lines reported by \citet{Kai04} towards TMC-1.
None of these lines coincide with the predicted strongest
lines of propylene.

The abundance of propylene relative to H$_2$ in TMC-1 is 4
$\times$ 10$^{-9}$, assuming that $N$(H$_2$) = 10$^{22}$ cm$^{-2}$ 
\citep{Cer87}. In spite of being a fairly
abundant hydrocarbon, it has not been previously considered in
chemical models of dark clouds. Reaction networks used to model
interstellar chemistry are usually tested by comparing with the
molecular abundances observed in TMC-1 \citep{Smi04,Woo07}. Thus,
much effort has been done to explore and include reactions leading
to the formation of the observed molecules, the abundances of
which are in general satisfactorily explained, although the
chemistry of not observed molecules is far from complete.

We have searched in the literature for reactions relevant to the
chemistry of propylene and performed chemical models to
investigate the most likely formation pathways. The synthesis of
many hydrocarbons in dark clouds involve a sequence of
ion-molecule reactions with the dissociative recombination (DR) of
a molecular cation and an electron as the last step \citep{Her89}.
Thus, the DR of the C$_3$H$_7^+$ ion, which is known to produce
C$_3$H$_6$ + H with a branching ratio of 0.42 \citep{Ehl03}, could
be a key step in the synthesis of propylene. If so, the problem of
producing propylene translates into how to form the ion
C$_3$H$_7^+$. Condensation reactions of an ionic and a neutral
hydrocarbon such as C$_2$H$_4^+$ + CH$_4$ $\rightarrow$
C$_3$H$_7^+$ + H as well as other similar ones, are endothermic,
too slow or tend to produce products other than C$_3$H$_7^+$
\citep{Ani03}. Consideration of radiative association reactions of
hydrocarbon ions with H$_2$ is an interesting possibility when
other channels are endothermic. We have checked that such a
possibility occurs for C$_2$H$_2^+$ and larger hydrocarbon ions
with 2, 3, 4 and 5 carbon atoms. However, the corresponding
reaction rates are not available in the literature and different
reaction channels may occur. Introducing these reactions with
realistic rate coefficient values and subsequent dissociative
recombination allows to build propylene at a reasonable level.

Propylene could also be one of the products in the DR of ions
larger than C$_3$H$_7^+$. Recent storage ring experiments have
shown that substantial fragmentation, with break of C-C bonds,
occurs in the DR of hydrocarbon ions \citep{Vig05}. For example,
chemical models predict an efficient synthesis of the ion
C$_6$H$_7^+$ in dark clouds, the DR of which is assumed to give
benzene and C$_6$H$_2$ (both predicted with an abundance of 
$\sim$ 10$^{-9}$; \citealt{Mce99}), although it could also
produce smaller neutral fragments such as propylene. The synthesis
of propylene through neutral-neutral reactions such as CH$_3$ +
C$_2$H$_4$ or C$_2$H$_3$ + CH$_4$ is not effective as these
reactions are endothermic.

Lastly, highly saturated hydrocarbons form easily in grain
surfaces by direct hydrogenation, although it is not clear how
these mantle-species would be desorbed to the gas phase at the low
temperatures of dark clouds. 
A possibility is to involve shocks and their role in the release 
of saturated adsorbed species from the surface of the grains 
\citep{Mar00}, or involve a local enhancement of the cosmic 
ionization rate which will have an important effect on the 
chemistry and evolutionary status of TMC-1 \citep{Har01}. 
However, the detected column density of methanol in TMC-1 
does not support this scenario. 
Furthermore, if this mechanism is working 
in TMC-1, other hot-core like molecules (CH$_3$OCH$_3$, 
CH$_3$OCOH, \ldots) should be detectectable there.
Recent work on non thermal desorption mechanisms from dust 
grains \citep{Gar07} show that some molecules formed on 
grain surfaces could go to the gas phase. The presence 
of hydrocarbons such as propene is an important information 
to improve the models as the hydrogenation proceeds on grain 
surfaces and the degree of surface hydrogenation increases with time.
In any case, it seems that gas-phase chemistry is missing important reactions 
involving small hydrocarbons, which could make molecules like 
propylene to have large abundances.
The detection of propylene suggest that other small hydrocarbons, 
like propane, could be present in dark clouds. A search for it is 
currently undergoing with the IRAM 30-m telescope.

\acknowledgments
We thank Spanish MEC for funding support through grants AYA2003-2785, 
AYA2006-14876, ESP2004-665, and AP2003-4619 (MA), and by DGU of the Madrid 
community government under IV-PRICIT project S-0505/ESP-0237 (ASTROCAM).
We also thank our referee for useful comments and suggestions.

\clearpage


\begin{deluxetable}{rcccccccc}
\tablecaption{Line parameters of the observed transitions from
gaussian fits\label{tab:lines}}
\tablehead{
\multicolumn{1}{c}{Rest Frequency \tablenotemark{a}} & Species & Transition & $E_{up}$ & $\mu^2S$ & $\int T_{\rm MB} dv$ & $V_{\rm LSR}$ & $\Delta v$ & $T_{\rm MB}$ \\
\multicolumn{1}{c}{(MHz)}     &         &            & (K)      & (D$^2$)    & (K km s$^{-1}$)     & (km s$^{-1}$) & (km s$^{-1}$) & (K)       \\
}
\tablecolumns{9}
\startdata
 84151.670 (7) & $A$    & $5_{15}-4_{14}$ & 13.9 & 0.610   & 0.014 (4)           & 5.72          & 0.46          & 0.028 (4) \\
 86651.566 (7) & $A$    & $5_{05}-4_{04}$ & 12.5 & 0.635   & 0.027 (4)           & 5.52          & 0.62          & 0.041 (4) \\
 87134.576 (6) & $A$    & $5_{24}-4_{23}$ & 19.8 & 0.534   & 0.005 (2)           & 5.68          & 0.32          & 0.016 (3) \\
 87677.584 (6) & $A$    & $5_{23}-4_{22}$ & 19.8 & 0.534   & 0.007 (2)           & 5.65          & 0.41          & 0.017 (3) \\
 89998.180 (7) & $A$    & $5_{14}-4_{13}$ & 14.8 & 0.610   & 0.012 (2)           & 5.53          & 0.31          & 0.036 (3) \\
100911.394 (8) & $A$    & $6_{16}-5_{15}$ & 18.8 & 0.741   & 0.010 (2)           & 5.62          & 0.43          & 0.022 (4) \\
103689.979 (8) & $A$    & $6_{06}-5_{05}$ & 17.5 & 0.761   & 0.020 (2)           & 5.71          & 0.55          & 0.033 (4) \\
\multicolumn{9}{c}{} \\
 86650.873 (24) & $E$    & $5_{05}-4_{04}$ & 12.5 & 0.635   & 0.017 (2)           & 5.68          & 0.35          & 0.044 (4) \\
 87137.941 ( 6) & $E$    & $5_{24}-4_{23}$ & 19.8 & 0.531   & 0.011 (2)           & 5.76          & 0.42          & 0.024 (3) \\
 87673.663 (29) & $E$    & $5_{23}-4_{22}$ & 19.8 & 0.531   & 0.008 (2)           & 5.68          & 0.42          & 0.018 (2) \\
 89997.232 (13) & $E$    & $5_{14}-4_{13}$ & 14.8 & 0.610   & 0.014 (2)           & 5.63          & 0.46          & 0.029 (3) \\
100911.141 ( 5) & $E$    & $6_{16}-5_{15}$ & 18.8 & 0.741   & 0.018 (2)           & 5.72          & 0.56          & 0.031 (4) \\
103689.117 (35) & $E$    & $6_{06}-5_{05}$ & 17.5 & 0.761   & 0.017 (2)           & 5.79          & 0.41          & 0.040 (4) \\
\enddata
\tablecomments{Number in parentheses are 1$\sigma$ uncertainties
in units of the last digits.}
\tablenotetext{a}{Frequencies for the $A$ especies are from \citet{Cer00} catalog, and those for the $E$ especies are the ones measured by \citet{Pea94}.}
\end{deluxetable}

\clearpage

\begin{deluxetable}{lcc|lcc}
\tablecaption{Column densities of several hydrocarbons in TMC-1
\label{tab:dens}}
\tablehead{
Species & $\mu$\tablenotemark{a}  & $N$  & Species & $\mu$\tablenotemark{a}  & $N$   \\
 &  (D) &  (cm$^{-2}$) &  &  (D) &  (cm$^{-2}$)  \\
}
\tablecolumns{6}
\startdata
C$_2$H $^{(1)}$      & 0.77  & 7$\times10^{13}$    & $c$-C$_3$H$_2$ $^{(2)}$ & 3.43  & 6$\times10^{13}$     \\
$l$-C$_3$H $^{(2)}$  & 3.55  & 8$\times10^{11}$    & H$_2$C$_4$ $^{(4)}$     & 4.10  & 8$\times10^{12}$      \\
$c$-C$_3$H $^{(2)}$  & 2.40  & 1$\times10^{13}$    & H$_2$C$_6$ $^{(5)}$     & 6.20  & 5$\times10^{11}$      \\
C$_4$H $^{(3)}$      & 0.87  & 3$\times10^{14}$    & CH$_3$CCH $^{(1)}$      & 0.78  & 8$\times10^{13}$      \\
C$_5$H $^{(4)}$      & 4.88  & 9$\times10^{12}$    & CH$_3$C$_4$H $^{(4)}$   & 1.21  & 1$\times10^{13}$        \\
C$_6$H $^{(2)}$      & 5.54  & 8$\times10^{12}$    & CH$_3$C$_6$H $^{(6)}$   & 1.50  & 3$\times10^{12}$ \\
$l$-C$_3$H$_2$ $^{(2)}$ & 4.10  & 2$\times10^{12}$ & CH$_2$CHCH$_3$ $^{(7)}$ & 0.36  & 4$\times10^{13}$ \\
\enddata
\tablenotetext{a}{Dipole moments are from JPL and CDMS catalogs}
\tablerefs{(1) \citet{Pra97}; (2) \citet{Fos01}; (3) \citet{Gue82}; (4) \citet{Ohi98}; (5) \citet{Lan97}; (6) \citet{Rem06}; (7) This work}
\end{deluxetable}

\clearpage


\clearpage

\begin{thebibliography}{}
\bibitem[Anicich(2003)]{Ani03}Anicich, V., G. 2003, JPL Publication 03-19
\bibitem[Cernicharo (1985)]{Cer85} Cernicharo, J. 1985, IRAM report No. 52
\bibitem[Cernicharo \& Gu\'elin (1987)]{Cer87} Cernicharo, J., Gu\'elin, M., 1987, \aap, 176, 299
\bibitem[Cernicharo et al. (2000)]{Cer00} Cernicharo, J., Gu\'elin, M., \& Kahane, C. 2000, \aaps, 142, 181
\bibitem[Cummins et al. (1986)]{Cum86} Cummins, S. E., Linke, R. A., \& Thaddeus, P. 1986, \apjs, 60, 819
\bibitem[Dickens et al. (2001)]{Dic01} Dickens, J. E., Langer, W. D., \& Velusamy, T. 2001, \apj, 558, 693
\bibitem[Ehlerding et al.(2003)]{Ehl03} Ehlerding, A., et al. 2003, \jpca, 107, 2179
\bibitem[Foss\'e et al. (2001)]{Fos01} Foss\'e, D., Cernicharo, J., Gerin, M., Cox, P., 2001, \apj, 552, 168
\bibitem[Friedel et al. (2004)]{Fri04} Friedel, D. N., Snyder, L. E., Turner, B. E., \& Remijan, A. 2004, \apj, 600, 234
\bibitem[Garrod et al. (2007)]{Gar07} Garrod, R. T., Wakelam, V., \& Herbst, E. 2007, \aap, 467, 1103
\bibitem[Gu\'elin et al. (1982)]{Gue82} Gu\'elin, M., Mezaoui, A., \& Friberg, P. 1982, \aap, 109, 23
\bibitem[Hartquist et al. (2001)]{Har01} Hartquist, T. W., Williams, D. A., \& Viti, S. 2001, \aap, 369, 605
\bibitem[Herbst \& Leung(1989)]{Her89} Herbst, E., \& Leung, C. M. 1989, \apjs, 69, 271
\bibitem[Kaifu et al. (2004)]{Kai04} Kaifu, N., Ohishi, M., Kawaguchi, K., et al., 2004, \pasj, 56, 69
\bibitem[Kalenskii et al. (2004)]{Kal04} Kalenskii, S. V., Slysh, V. I., Goldsmith, P. F., \& Johansson, L. E. B. 2004, \apj, 610, 329
\bibitem[Langer et al. (1997)]{Lan97} Langer, W. D., Velusamy, T., et al. 1997, \apjl, 480, L63
\bibitem[Lide \& Mann (1957)]{Lid57} Lide, D. R., Mann, D. E., 1957, J. Chem. Phys., 27, 868
\bibitem[Markwick et al. (2000)]{Mar00} Markwick, A. J., Millar, T. J., \& Charnley, S. B. 2000, \apj, 535, 256
\bibitem[McEwan et al.(1999)]{Mce99} McEwan, M. J., Scott, G. B. I., Adams, N. G., et al. 1999, \apj, 513, 287
\bibitem[M\"uller et al. (2001)]{Mul01} M\"uller, H. S. P., Thorwirth, S., Roth, D. A., \& Winnewisser, G. 2001, \aap, 370, L49
\bibitem[M\"uller et al. (2005)]{Mul05} M\"uller, H. S. P., Schl\"oder, F., Stutzki, J., \& Winnewisser, G. 2005, J. Mol. Struct., 742, 215
\bibitem[Nummelin et al. (1998)]{Num98} Nummelin, A., Bergman, P., Hjalmarson, \AA., et al. 1998, \apjs, 117, 427
\bibitem[Ohishi \& Kaifu (1998)]{Ohi98} Ohishi, M., Kaifu, N., 1998, \fad, 109, 205
\bibitem[Pearson et al. (1994)]{Pea94} Pearson, J. C., Sastry, K. V. L. N., Herbst, E., \& De Lucia, F. C. 1994, \jmsp, 166, 120
\bibitem[Peng et al. (1998)]{Pen98} Peng, R., Langer, W. D., Velusamy, T., Kuiper, T. B. H., \& Levin, S. 1998, \apj, 497, 842
\bibitem[Pickett et al. (1998)]{Pic98} Pickett, H. M., Poynter, R. L., Cohen, E. A., et al. 1998, \jqsrt, 60, 883
\bibitem[Pratap et al. (1997)]{Pra97} Pratap, P., Dickens, J. E., Snell, R. L., et al. 1997, \apj, 486, 862
\bibitem[Remijan et al. (2006)]{Rem06} Remijan, A. J., Hollis, J. M., Snyder, L. E., Jewell, P. R., \& Lovas, F. J. 2006, \apjl, 643, L37
\bibitem[Schilke et al. (1997)]{Sch97} Schilke, P., Groesbeck, T. D., Blake, G. A., \& Phillips, T. G. 1997, \apjs, 108, 301
\bibitem[Smith et al.(2004)]{Smi04} Smith, I. W. M., Herbst, E., \& Chang, Q. 2004, \mnras, 350, 323
\bibitem[Snyder et al. (2006)]{Sny06} Snyder, L. E., Hollis, J. M., Jewell, P. R., Lovas, F. J., \& Remijan, A. 2006, \apj, 647, 412
\bibitem[Sutton et al. (1985)]{Sut85} Sutton, E. C., Blake, G. A., Masson, C. R., \& Phillips, T. G. 1985, \apjs, 58, 341
\bibitem[Thum et al. (1995)]{Thu95} Thum, C., Sievers, A., Navarro, S., Brunswig, W., \& Pe\~nalver, J. 1995, IRAM report No. 228
\bibitem[Turner (1989)]{Tur89} Turner, B. E. 1989, \apjs, 70, 539
\bibitem[Viggiano et al.(2005)]{Vig05} Viggiano, A. A., Ehlerding, A., Arnold, S. T., \& Larsson, M. 2005, \jpcs, 4, 191
\bibitem[Wlodarczak et al. (1994)]{Wlo94} Wlodarczak, G., Demaison, J., Heineking, N., Cs\'asz\'ar, G., 1994, \jmsp, 167, 239
\bibitem[Woodall et al.(2007)]{Woo07} Woodall, J., Ag\'undez, M., Markwick-Kemper, A. J., \& Millar, T. J. 2007, \aap, 466, 1197
\end{thebibliography}
\end{document}